\documentclass[aps,prl,twocolumn,superscriptaddress]{revtex4-1}

\usepackage{amssymb}
\usepackage{amsmath}
\usepackage{graphicx}
\usepackage{dcolumn}
\usepackage{bm}
\usepackage{color}

\begin{document}

\title{Anisotropic magneto-capacitance in ferromagnetic-plate capacitors}

\author{J.~A.~Haigh}
\affiliation{Hitachi Cambridge Laboratory, J.~J.~Thomson Ave., Cambridge CB3 0HE, United Kingdom}
\author{C.~Ciccarelli}
\affiliation{Microelectronics Group, Cavendish Laboratory, University of Cambridge, J.~J.~Thomson Ave., Cambridge CB3 0HE, United Kingdom}
\author{A.~C.~Betz}
\affiliation{Hitachi Cambridge Laboratory, J.~J.~Thomson Ave., Cambridge CB3 0HE, United Kingdom}
\author{A.~Irvine}
\affiliation{Microelectronics Group, Cavendish Laboratory, University of Cambridge, J.~J.~Thomson Ave., Cambridge CB3 0HE, United Kingdom}
\author{V.~Nov\'{a}k}
\affiliation{Institute of Physics ASCR, v.v.i., Cukrovarnick\'{a} 10, 16253 Praha 6, Czech Republic}
\author{T.~Jungwirth}
\affiliation{Institute of Physics ASCR, v.v.i., Cukrovarnick\'{a} 10, 16253 Praha 6, Czech Republic}
\affiliation{School of Physics and Astronomy, University of Nottingham, Nottingham NG7 2RD, UK}
\author{J.~Wunderlich}
\affiliation{Hitachi Cambridge Laboratory, J.~J.~Thomson Ave., Cambridge CB3 0HE, United Kingdom}
\affiliation{Institute of Physics ASCR, v.v.i., Cukrovarnick\'{a} 10, 16253 Praha 6, Czech Republic}
\date{\today}
\begin{abstract}

The capacitance of a parallel plate capacitor can depend on applied magnetic field. Previous studies have identified capacitance changes induced via classical Lorentz force or spin-dependent Zeeman effects. Here we measure a magnetization direction dependent capacitance in parallel-plate capacitors where one plate is a ferromagnetic semiconductor, gallium manganese arsenide. This anisotropic magneto-capacitance is due to the anisotropy in the density of states dependent on the magnetization through the strong spin-orbit interaction.

\end{abstract}

\maketitle

Capacitance, the ability of a body to retain charge is defined by the relation $1/C_g=dV_e/dq$, the ratio of the change in electrostatic potential to the amount of charge added. In a simple parallel plate capacitor one normally calculates this capacitance through the change in electrostatic potential by integrating the electric field $E$ due to charges on the surface of two metallic plates over the separating distance $d$. However, corrections to this electrostatic picture can be important, and other contributions to the change in potential when additional charge is added must be taken into account. It is often helpful to reformulate these corrections as effective series capacitances in series with a geometrical capacitance expected from the classical picture \cite{stern_self-consistent_1972,smith_direct_1985,buttiker_mesoscopic_1993}. In particular, the effect of change in chemical potential due to the finite density of states can be important in low dimensional systems. This has been exploited for example in two-dimensional electron gases where the additional chemical contribution to the potential, the electron compressibility, allows probing of Landau levels in the density of states in the quantum Hall regime \cite{smith_direct_1985}.

In this Letter, we exploit this kind of capacitance correction to demonstrate an anisotropic magneto-capacitance (AMC). This is analogous to anisotropic magneto-resistance (AMR) \cite{thomson_electro-dynamic_1856}, an important technology in magnetic field sensing applications \cite{thompson_thin_1975} and of a similarly relativistic magnetic origin, but in this different fundamental electrical circuit element.

In general, magnetic effects on transport properties such as AMR can be ascribed to three different categories: ordinary (orbital), due to the Lorentz force; spin-dependent, due to splitting of spin sub-bands through ferromagnetism or the Zeeman effect; and extraordinary, relativistic in origin through the spin-orbit interaction. Some well known examples of these effects in resistance are Lorentz magneto-resistance, giant magneto-resistance (GMR) \cite{Baibich_giant_1988}, and AMR \cite{mcguire_anisotropic_1975} respectively. Classifying magneto-capacitance along similar lines, both ordinary and spin-dependent effects have been observed previously. Changes in capacitance as a function of in-plane magnetic field have been measured in two-dimensional electron gases and attributed to combined Lorentz force and quantum confinement effects \cite{hampton_capacitance_1995,jungwirth_capacitance_1995}; spin dependent effects have been considered theoretically \cite{zhang_spin-dependent_1999}, and experimentally measured due to the Zeeman splitting in Pd plate capacitors \cite{mccarthy_magnetocapacitance:_2003} and in magnetic tunnel junctions several measurements have shown changes in capacitance as a function of relative magnetization orientation \cite{Kaiju_magnetocapacitance_2002,padhan_frequency-dependent_2007,chang_extraction_2010}. The demonstration in this paper is of the third class, an anisotropic magneto-capacitance whose origin is in the spin-orbit interaction.

\begin{figure}%
\includegraphics[width=0.95\columnwidth]{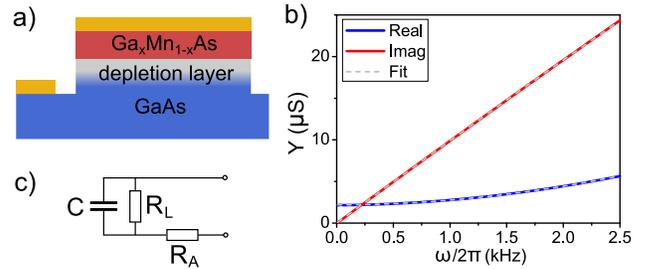}%
\caption{a) device structure. b) example measurement and fit of admittance of p-n junction capacitor. The leakage resistance and the access resistance are evident in the offset and frequency dependent of the real part respectively. c) lumped element circuit diagram used for fitting.}%
\label{fig1}%
\end{figure}

It is also worth noting the connection of a modulation in capacitance to that of a different quantity, the chemical potential. Changes in chemical potential can be measured by exploiting the capacitance of a structure and measuring the rearrangement of charge to maintain equilibrium. By contrast, a change in capacitance does not change the equilibrium charge distribution, but the variation of charge with respect to voltage (the definition of capacitance). While magnetically induced changes in chemical potential have been measured experimentally in single electron transistor devices as spin-dependent magneto-coulomb oscillations \cite{ono_enhanced_1997,ono_ferromagnetic_1998}, and spin-orbit interaction induced coulomb-blockade AMR \cite{wunderlich_coulomb_2006,schlapps_transport_2009,bernand-mantel_anisotropic_2009,ciccarelli_spin_2012}, anisotropies in the capacitance were not observed or considered. There, the capacitance of the devices was small and remains constant so that the sensitivity is to the chemical potential; if a change in the capacitance were observed in coulomb blockaded transport it would have appeared in the period of the oscillating conductivity, rather than the observed shift.


The system in which we chose to demonstrate AMC is the ferromagnetic semiconductor (Ga,Mn)As \cite{ohno_gamnas:_1996}. This material is known to exhibit anisotropy in the density of states \cite{jungwirth_theory_2006}, as is experimentally demonstrated by the large tunneling anisotropic magneto-resistance (TAMR) \cite{gould_tunneling_2004}. The 25\,nm (Ga,Mn)As layer is grown by standard low temperature molecular beam epitaxy, with manganese concentration of $8\%$, on a 1.5\,$\mu$m $n$-GaAs layer (nominal doping $n\approx1\times10^{17}$\,cm$^{-3}$) with a contact layer $n^{+}\approx7\times10^{18}$\,cm$^{-3}$ below. The depletion layer of this p-n junction is used as the dielectric of the capacitor. Wafers are fabricated into large-area planar structures using optical lithography and wet chemical etching and contacts to the (Ga,Mn)As and n$^{+}$-GaAs layers are made through thermal evaporation of Cr/Au and AuGeNi respectively (fig. \ref{fig1}(a)). The (Ga,Mn)As is as-grown with $T_c$ around 42\,K.

Experiments are performed in a 3-axis vector magnet cryostat. A lock-in amplifier is used to measure the quadrature components of the current with a small applied voltage excitation across the capacitor as a function of excitation frequency from $\sim$dc to 2.5\,kHz. An example of such a measurement is shown in fig. \ref{fig1}(b). A linear dependence in the imaginary (out-of-phase) part of the admittance on excitation frequency indicates the dominance of capacitance in this device. In addition, in the real (in-phase) admittance there is a offset due to a parallel ohmic leakage resistance through the dielectric layer and a quadratic part due to this leakage and the resistance of the leads. To extract the capacitance the complex admittance is fitted to a simple lumped element circuit model (fig. \ref{fig1}(c)) which includes the capacitance $C$, leakage resistance $R_L$, and an access resistance $R_A$ in the $n$-GaAs contact layer, which gives:
\begin{multline}
Y = \frac{1}{\left(1+\frac{R_A}{R_L}\right)^2 + \omega^2 C^2 R_A^2} \\ \left( i \omega C + \frac{1}{R_L}\left(1+\frac{R_A}{R_L}\right) + \omega^2 R_A^2 C^2\right).
\label{eq:}
\end{multline}
The prefactor has only a small effect as $R_A/R_L$ is small and we work at low frequencies. The quality of the fit is excellent as can be seen from the dashed lines in fig \ref{fig1}(b). This measurement is repeated as a function of magnetic field direction in order to rotate the magnetization in the (Ga,Mn)As.

\begin{figure}%
\includegraphics[width=0.8\columnwidth]{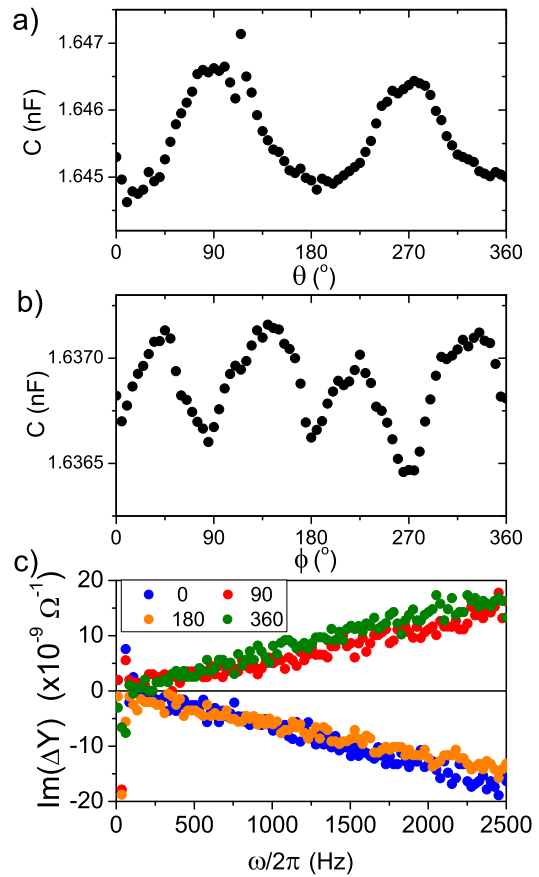}%
\caption{2 K measurements of capacitance for rotations of the magnetization (a) out-of plane and (b)  in plane with a 1\,T magnetic field. The capacitance is extracted from the fitting the complex admittance, as discussed in text, and a small temperature dependent drift has been subtracted. (c) The difference in the imaginary part of the admittance compared to the mean as a function of frequency for various angles $\theta$ of applied field with respect to the plane (same data as corresponding angles in (a)).}%
\label{rot}%
\end{figure}

In fig. \ref{rot} the capacitance is plotted as a function of in-plane (a) and out-of-plane (b) magnetic field orientation of magnitude 0.5\,T. There is a clear cubic in-plane symmetry, and uniaxial out-of-plane symmetry to the change in capacitance. This is the symmetry that might be expected due to the bulk cubic crystal symmetry in the plane and uniaxial out-of-plane anisotropy due to the compressive strain on the (Ga,Mn)As epilayer from the GaAs substrate.. The size of the modulation of the capacitance is $0.1\%$ out-of-plane and $0.03\%$ in-plane. This small ratio could in principle be strengthened by increasing the geometrical capacitance, as will be discussed in the analysis. For these rotation experiments, effectively no change in the leakage or access resistance is measured within experimental accuracy. To conclusively demonstrate that it is a change in capacitance which we are measuring, we also plot in fig. \ref{rot}(c) the difference in the imaginary part of the admittance from the mean for four directions of the magnetization with respect to the plane. As expected, these are straight lines as a function of frequency given by $\Im(\Delta Y) \approx \omega \Delta C$. 

\begin{figure}%
\includegraphics[width=\columnwidth]{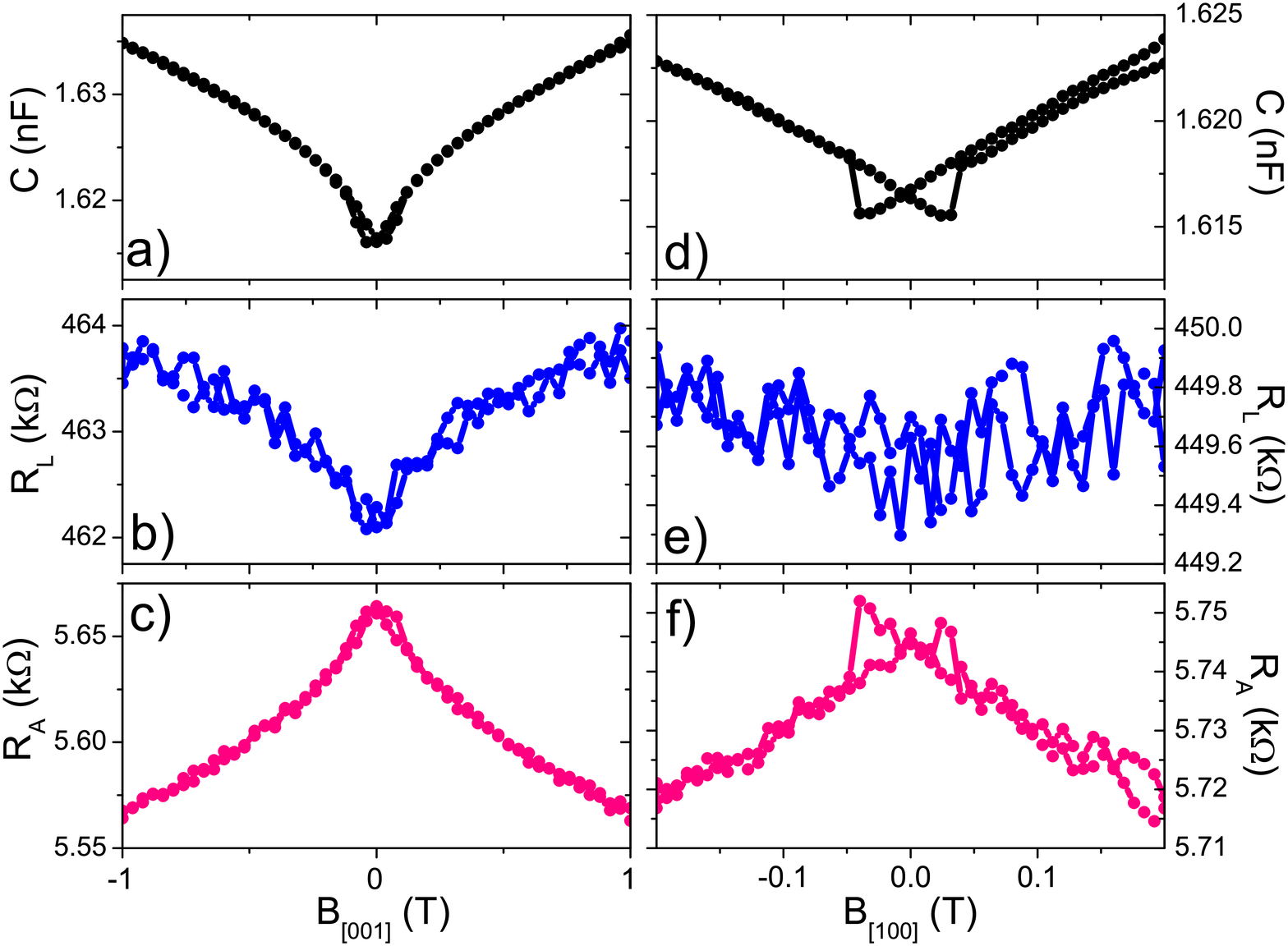}%
\caption{2 K  hysteresis loops for out-of plane (a)-(c) and in plane (d)-(e) magnetic field directions. The extracted capacitance (a),(d), leakage resistance (b),(e) and access resistance (c),(f) are plotted as a function of magnetic field.}%
\label{hyst}%
\end{figure}

The capacitance is also measured as a function of magnetic field strength (fig. \ref{hyst}). Here, hysteresis is observed for in-plane field directions evidencing the magnetic origin of this effect. An isotropic linear magneto-capacitance is also observed extending to high magnetic fields where the magnetisation is already saturated. A similiar effect has been seen in (Ga,Mn)As in CBAMR, and was understood through two mechanism \cite{ciccarelli_spin_2012}. Firstly the field dependence of a small amount of unsaturated manganese moments, and secondly the Zeeman splitting of the bands. While we are sensitive to a different quantity in our experiments, both these mechanisms likely play a role in the measured linear magneto-capacitance, as well as isotropic magneto-capacitance contributions in the non-magnetic n-GaAs \cite{tongay_magnetodielectric_2009}. In addition, the access and leakage resistance both show separately a small magneto-resistance. We note here that while it has been shown that a series capacitance can mean that any magneto-resistance appears as a spurious measured magneto-capacitance \cite{catalan_magnetocapacitance_2006}, this can be excluded in our devices. In order for the observed magneto-resistance to appear in the measured capacitance, it would require that the series capacitance be in excess of 100\,nF, much greater than can be reasonable expected given the geometry of our devices.

There are two possible contributions to the AMC that we observe. One is related to the modulation of the density of states and the other to that of the chemical potential. The latter is specific to the p-n junction structure of these devices, but both are extra-ordinary magneto-capacitance effects and we attempt to differentiate the two by treating them separately.

The effect of the density of states on capacitance is often evident in low dimensional systems where the density of states is small and the additional change in potential $\Delta \mu /e$ is inversely proportional to the density of states. In 2D systems, for example, this is often reformulated as an additional capacitance per area $C_q/A = e^2\rho_{2D}$, for density of states $\rho_{\text{2D}}$, in series with the electro-static geometric capacitance. This is termed the quantum capacitance \cite{Iafrate_capacitive_1995} in the zero temperature limit and electron compressibility \cite{eisenstein_compressibility_1994} when the density of states is smeared out at finite temperature. The effect has important consequences in capacitors formed from 2D systems such as semiconductor hetereostructures \cite{smith_two-dimensional_1986} and graphene \cite{Ponomarenko_density_2010}, 1D systems such as carbon nanotubes \cite{Ilani_measurement_2006} and 0D systems such as quantum dots \cite{hansen_zeeman_1989,ashoori_single-electron_1992}, where changes in the density of states have been measured through the effect on the measured capacitance. These measurements were possible through the ability of some external parameter to change the density of states, which then allows the separation of the contributions to capacitance which would be otherwise indistinguishable. In our experiment, this handle is provided by the direction of the magnetization with respect to the crystallographic axes.

In 3D systems, the density of states is generally much larger and the corrections to the capacitance can often be neglected. In addition, the separation of the plates becomes an ill-defined quantity once the finite screening length in the plate is taken into account \cite{buttiker_capacitance_1993}. The chemical potential contribution to the capacitance then depends on how the charge is distributed in the contact. This distribution of charge is itself defined by the competition between the electrostatic and potential energies. However, based on the Thomas-Fermi screening length one can formulate an effective series capacitance formulated in 3D \cite{buttiker_capacitance_1993}, with an equivalent dependence on the density of states \cite{kopp_calculation_2009} similar to that for the electron compressibility, the kinetic capacitance.
\begin{equation}
C_{k}/A = \sqrt{e^2 \epsilon_0\epsilon_{eff} \rho_{\text{3D}} }
\label{eq:ck}
\end{equation}
Where $\rho_{\text{3D}}$ is the density of states and $\epsilon_0\epsilon_{\text{eff}}$ is the effective dielectric constant of the contact material, which we take to be that of bulk GaAs. To asses whether the change in resistance can be explained by the anisotropy of the density of states in the ferromagnet, we use an approximated value of the total density of states to obtain an estimate of the corresponding change which would be needed to give the observed capacitance modulation. Taking $\rho_{\text{3D}}\approx 10^{46}$\,J$^{-1}$m$^{-3}$ \cite{neumaier_all-electrical_2009}, we estimate $C_k/A\approx$50\,fF$\mu$m$^{-2}$. The change in the total capacitance of the device $C_T$ can be related to the change in the kinetic capacitance by
\begin{eqnarray}
\frac{\Delta C_T}{C_T} = \frac{C_T}{C_k}\frac{\Delta C_k}{C_k}.
\label{eq:c}
\end{eqnarray}
It can be seen from this equation that increasing the ratio of total to kinetic capacitance would increase the size of the modulation. The total capacitance per area of our device is $\approx$2\,fF$\mu$m$^{-2}$. From these values a $\approx$5\% change in the density of states would be needed to explain the 0.1\% change in capacitance measured for the out-of-plane field rotation, a value which is reasonable in these as-grown material.

\begin{figure}%
\includegraphics[width=0.8\columnwidth]{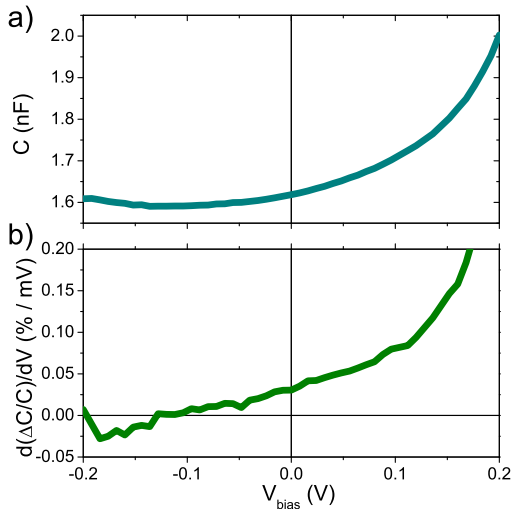}%
\caption{a) capacitance-voltage measurements at 2 K on p-n junction device. b) change in capacitance as a function of bias.}%
\label{cv}%
\end{figure}

We now consider the effect of the anisotropy in chemical potential. Because our capacitors are p-n junctions, any change in the difference in chemical potential across the device would change the width of the depletion layer in the $n$-GaAs and thus modulate the capacitance in a different way. To ascertain whether this is possible, we measure the capacitance of our devices as a function of dc bias voltage. We can then compare the voltage which is needed to obtain a comparable shift in the capacitance to that expected from the anisotropy of the chemical potential (fig. \ref{cv}). To obtain the same change in capacitance we observe for the out-of-plane rotation the change in chemical potential would need to be $\approx$5\,mV. This value is much larger than that measured in (Ga,Mn)As through CBAMR in both disordered planar devices \cite{wunderlich_coulomb_2006} and aluminum single electron transistors \cite{ciccarelli_spin_2012}. In this case we can largely assign the observed AMC to a density of states effect.

As the density of states AMC is not reliant on the p-n junction structure we have used, it should also be observable more generally in normal dielectric capacitors with ferromagnetic, or indeed antiferromagnetic contacts with strong spin-orbit coupling. Conditions are more favorable in materials with low absolute density of states, as this reduces the total kinetic capacitance and therefore makes it easier to achieve a comparable geometric capacitance with typical thicknesses and strengths of dielectric materials. In this light, antiferromagnetic semiconductors would be a prime candidate for room temperature devices with large effects. In addition, the magneto-capacitance observed by McCarthy et al. \cite{mccarthy_magnetocapacitance:_2003} in metal plate capacitors provides a hint that AMC could be observed in metallic ferromagnets, where a possible candidate might be CoPt whose strong spin-orbit interaction, and the resulting considerable density of states anisotropies, facilitates a large TAMR \cite{shick_prospect_2006,park_tunneling_2008}.

We would like to acknowledge support from Hitachi Cambridge Laboratory, the EU European Research Council (ERC) Advanced Grant No. 268066, the Ministry of Education of the Czech Republic Grant No.LM2011026, and the Grant Agency of the Czech Republic Grant No. 14-37427G.

\bibliography{bibliography}{}

\begin{thebibliography}{37}
\expandafter\ifx\csname natexlab\endcsname\relax\def\natexlab#1{#1}\fi
\expandafter\ifx\csname bibnamefont\endcsname\relax
  \def\bibnamefont#1{#1}\fi
\expandafter\ifx\csname bibfnamefont\endcsname\relax
  \def\bibfnamefont#1{#1}\fi
\expandafter\ifx\csname citenamefont\endcsname\relax
  \def\citenamefont#1{#1}\fi
\expandafter\ifx\csname url\endcsname\relax
  \def\url#1{\texttt{#1}}\fi
\expandafter\ifx\csname urlprefix\endcsname\relax\def\urlprefix{URL }\fi
\providecommand{\bibinfo}[2]{#2}
\providecommand{\eprint}[2][]{\url{#2}}

\bibitem[{\citenamefont{Stern}(1972)}]{stern_self-consistent_1972}
\bibinfo{author}{\bibfnamefont{F.}~\bibnamefont{Stern}},
  \bibinfo{journal}{Physical Review B} \textbf{\bibinfo{volume}{5}},
  \bibinfo{pages}{4891} (\bibinfo{year}{1972}).

\bibitem[{\citenamefont{Smith et~al.}(1985)\citenamefont{Smith, Goldberg,
  Stiles, and Heiblum}}]{smith_direct_1985}
\bibinfo{author}{\bibfnamefont{T.~P.} \bibnamefont{Smith}},
  \bibinfo{author}{\bibfnamefont{B.~B.} \bibnamefont{Goldberg}},
  \bibinfo{author}{\bibfnamefont{P.~J.} \bibnamefont{Stiles}},
  \bibnamefont{and} \bibinfo{author}{\bibfnamefont{M.}~\bibnamefont{Heiblum}},
  \bibinfo{journal}{Physical Review B} \textbf{\bibinfo{volume}{32}},
  \bibinfo{pages}{2696} (\bibinfo{year}{1985}).

\bibitem[{\citenamefont{B\"uttiker et~al.}(1993)\citenamefont{B\"uttiker,
  Thomas, and Pr\^etre}}]{buttiker_mesoscopic_1993}
\bibinfo{author}{\bibfnamefont{M.}~\bibnamefont{B\"uttiker}},
  \bibinfo{author}{\bibfnamefont{H.}~\bibnamefont{Thomas}}, \bibnamefont{and}
  \bibinfo{author}{\bibfnamefont{A.}~\bibnamefont{Pr\^etre}},
  \bibinfo{journal}{Physics Letters A} \textbf{\bibinfo{volume}{180}},
  \bibinfo{pages}{364} (\bibinfo{year}{1993}).

\bibitem[{\citenamefont{Thomson}(1856)}]{thomson_electro-dynamic_1856}
\bibinfo{author}{\bibfnamefont{W.}~\bibnamefont{Thomson}},
  \bibinfo{journal}{Proceedings of the Royal Society of London}
  \textbf{\bibinfo{volume}{8}}, \bibinfo{pages}{546} (\bibinfo{year}{1856}).

\bibitem[{\citenamefont{Thompson et~al.}(1975)\citenamefont{Thompson, Romankiw,
  and Mayadas}}]{thompson_thin_1975}
\bibinfo{author}{\bibfnamefont{D.~A.} \bibnamefont{Thompson}},
  \bibinfo{author}{\bibfnamefont{L.}~\bibnamefont{Romankiw}}, \bibnamefont{and}
  \bibinfo{author}{\bibfnamefont{A.}~\bibnamefont{Mayadas}},
  \bibinfo{journal}{{IEEE} Transactions on Magnetics}
  \textbf{\bibinfo{volume}{11}}, \bibinfo{pages}{1039} (\bibinfo{year}{1975}).

\bibitem[{\citenamefont{Baibich et~al.}(1988)\citenamefont{Baibich, Broto,
  Fert, Nguyen Van~Dau, Petroff, Etienne, Creuzet, Friederich, and
  Chazelas}}]{Baibich_giant_1988}
\bibinfo{author}{\bibfnamefont{M.~N.} \bibnamefont{Baibich}},
  \bibinfo{author}{\bibfnamefont{J.~M.} \bibnamefont{Broto}},
  \bibinfo{author}{\bibfnamefont{A.}~\bibnamefont{Fert}},
  \bibinfo{author}{\bibfnamefont{F.}~\bibnamefont{Nguyen Van~Dau}},
  \bibinfo{author}{\bibfnamefont{F.}~\bibnamefont{Petroff}},
  \bibinfo{author}{\bibfnamefont{P.}~\bibnamefont{Etienne}},
  \bibinfo{author}{\bibfnamefont{G.}~\bibnamefont{Creuzet}},
  \bibinfo{author}{\bibfnamefont{A.}~\bibnamefont{Friederich}},
  \bibnamefont{and} \bibinfo{author}{\bibfnamefont{J.}~\bibnamefont{Chazelas}},
  \bibinfo{journal}{Physical Review Letters} \textbf{\bibinfo{volume}{61}},
  \bibinfo{pages}{2472} (\bibinfo{year}{1988}).

\bibitem[{\citenamefont{McGuire and Potter}(1975)}]{mcguire_anisotropic_1975}
\bibinfo{author}{\bibfnamefont{T.}~\bibnamefont{McGuire}} \bibnamefont{and}
  \bibinfo{author}{\bibfnamefont{R.}~\bibnamefont{Potter}},
  \bibinfo{journal}{{IEEE} Transactions on Magnetics}
  \textbf{\bibinfo{volume}{11}}, \bibinfo{pages}{1018} (\bibinfo{year}{1975}).

\bibitem[{\citenamefont{Hampton et~al.}(1995)\citenamefont{Hampton, Eisenstein,
  Pfeiffer, and West}}]{hampton_capacitance_1995}
\bibinfo{author}{\bibfnamefont{J.}~\bibnamefont{Hampton}},
  \bibinfo{author}{\bibfnamefont{J.}~\bibnamefont{Eisenstein}},
  \bibinfo{author}{\bibfnamefont{L.}~\bibnamefont{Pfeiffer}}, \bibnamefont{and}
  \bibinfo{author}{\bibfnamefont{K.}~\bibnamefont{West}},
  \bibinfo{journal}{Solid State Communications} \textbf{\bibinfo{volume}{94}},
  \bibinfo{pages}{559} (\bibinfo{year}{1995}).

\bibitem[{\citenamefont{Jungwirth and
  Smr\v{c}ka}(1995)}]{jungwirth_capacitance_1995}
\bibinfo{author}{\bibfnamefont{T.}~\bibnamefont{Jungwirth}} \bibnamefont{and}
  \bibinfo{author}{\bibfnamefont{L.}~\bibnamefont{Smr\v{c}ka}},
  \bibinfo{journal}{Physical Review B} \textbf{\bibinfo{volume}{51}},
  \bibinfo{pages}{10181} (\bibinfo{year}{1995}).

\bibitem[{\citenamefont{Zhang}(1999)}]{zhang_spin-dependent_1999}
\bibinfo{author}{\bibfnamefont{S.}~\bibnamefont{Zhang}},
  \bibinfo{journal}{Physical Review Letters} \textbf{\bibinfo{volume}{83}},
  \bibinfo{pages}{640} (\bibinfo{year}{1999}).

\bibitem[{\citenamefont{McCarthy et~al.}(2003)\citenamefont{McCarthy, Hebard,
  and Arnason}}]{mccarthy_magnetocapacitance:_2003}
\bibinfo{author}{\bibfnamefont{K.~T.} \bibnamefont{McCarthy}},
  \bibinfo{author}{\bibfnamefont{A.~F.} \bibnamefont{Hebard}},
  \bibnamefont{and} \bibinfo{author}{\bibfnamefont{S.~B.}
  \bibnamefont{Arnason}}, \bibinfo{journal}{Physical Review Letters}
  \textbf{\bibinfo{volume}{90}}, \bibinfo{pages}{117201}
  (\bibinfo{year}{2003}).

\bibitem[{\citenamefont{Kaiju et~al.}(2002)\citenamefont{Kaiju, Fujita,
  Morozumi, and Shiiki}}]{Kaiju_magnetocapacitance_2002}
\bibinfo{author}{\bibfnamefont{H.}~\bibnamefont{Kaiju}},
  \bibinfo{author}{\bibfnamefont{S.}~\bibnamefont{Fujita}},
  \bibinfo{author}{\bibfnamefont{T.}~\bibnamefont{Morozumi}}, \bibnamefont{and}
  \bibinfo{author}{\bibfnamefont{K.}~\bibnamefont{Shiiki}},
  \bibinfo{journal}{Journal of Applied Physics} \textbf{\bibinfo{volume}{91}},
  \bibinfo{pages}{7430} (\bibinfo{year}{2002}).

\bibitem[{\citenamefont{Padhan et~al.}(2007)\citenamefont{Padhan, LeClair,
  Gupta, Tsunekawa, and Djayaprawira}}]{padhan_frequency-dependent_2007}
\bibinfo{author}{\bibfnamefont{P.}~\bibnamefont{Padhan}},
  \bibinfo{author}{\bibfnamefont{P.}~\bibnamefont{LeClair}},
  \bibinfo{author}{\bibfnamefont{A.}~\bibnamefont{Gupta}},
  \bibinfo{author}{\bibfnamefont{K.}~\bibnamefont{Tsunekawa}},
  \bibnamefont{and} \bibinfo{author}{\bibfnamefont{D.~D.}
  \bibnamefont{Djayaprawira}}, \bibinfo{journal}{Applied Physics Letters}
  \textbf{\bibinfo{volume}{90}}, \bibinfo{pages}{142105}
  (\bibinfo{year}{2007}).

\bibitem[{\citenamefont{Chang et~al.}(2010)\citenamefont{Chang, Li, Huang,
  Tung, Tong, and Lin}}]{chang_extraction_2010}
\bibinfo{author}{\bibfnamefont{Y.-M.} \bibnamefont{Chang}},
  \bibinfo{author}{\bibfnamefont{K.-S.} \bibnamefont{Li}},
  \bibinfo{author}{\bibfnamefont{H.}~\bibnamefont{Huang}},
  \bibinfo{author}{\bibfnamefont{M.-J.} \bibnamefont{Tung}},
  \bibinfo{author}{\bibfnamefont{S.-Y.} \bibnamefont{Tong}}, \bibnamefont{and}
  \bibinfo{author}{\bibfnamefont{M.-T.} \bibnamefont{Lin}},
  \bibinfo{journal}{Journal of Applied Physics} \textbf{\bibinfo{volume}{107}},
  \bibinfo{pages}{093904} (\bibinfo{year}{2010}).

\bibitem[{\citenamefont{Ono et~al.}(1997)\citenamefont{Ono, Shimada, and
  Ootuka}}]{ono_enhanced_1997}
\bibinfo{author}{\bibfnamefont{K.}~\bibnamefont{Ono}},
  \bibinfo{author}{\bibfnamefont{H.}~\bibnamefont{Shimada}}, \bibnamefont{and}
  \bibinfo{author}{\bibfnamefont{Y.}~\bibnamefont{Ootuka}},
  \bibinfo{journal}{Journal of the Physical Society of Japan}
  \textbf{\bibinfo{volume}{66}}, \bibinfo{pages}{1261} (\bibinfo{year}{1997}).

\bibitem[{\citenamefont{Ono et~al.}(1998)\citenamefont{Ono, Shimada, and
  Ootuka}}]{ono_ferromagnetic_1998}
\bibinfo{author}{\bibfnamefont{K.}~\bibnamefont{Ono}},
  \bibinfo{author}{\bibfnamefont{H.}~\bibnamefont{Shimada}}, \bibnamefont{and}
  \bibinfo{author}{\bibfnamefont{Y.}~\bibnamefont{Ootuka}},
  \bibinfo{journal}{Solid-State Electronics} \textbf{\bibinfo{volume}{42}},
  \bibinfo{pages}{1407} (\bibinfo{year}{1998}).

\bibitem[{\citenamefont{Wunderlich et~al.}(2006)\citenamefont{Wunderlich,
  Jungwirth, Kaestner, Irvine, Shick, Stone, Wang, Rana, Giddings, Foxon
  et~al.}}]{wunderlich_coulomb_2006}
\bibinfo{author}{\bibfnamefont{J.}~\bibnamefont{Wunderlich}},
  \bibinfo{author}{\bibfnamefont{T.}~\bibnamefont{Jungwirth}},
  \bibinfo{author}{\bibfnamefont{B.}~\bibnamefont{Kaestner}},
  \bibinfo{author}{\bibfnamefont{A.~C.} \bibnamefont{Irvine}},
  \bibinfo{author}{\bibfnamefont{A.~B.} \bibnamefont{Shick}},
  \bibinfo{author}{\bibfnamefont{N.}~\bibnamefont{Stone}},
  \bibinfo{author}{\bibfnamefont{K.-Y.} \bibnamefont{Wang}},
  \bibinfo{author}{\bibfnamefont{U.}~\bibnamefont{Rana}},
  \bibinfo{author}{\bibfnamefont{A.~D.} \bibnamefont{Giddings}},
  \bibinfo{author}{\bibfnamefont{C.~T.} \bibnamefont{Foxon}},
  \bibnamefont{et~al.}, \bibinfo{journal}{Physical Review Letters}
  \textbf{\bibinfo{volume}{97}}, \bibinfo{pages}{077201}
  (\bibinfo{year}{2006}).

\bibitem[{\citenamefont{Schlapps et~al.}(2009)\citenamefont{Schlapps, Lermer,
  Geissler, Neumaier, Sadowski, Schuh, Wegscheider, and
  Weiss}}]{schlapps_transport_2009}
\bibinfo{author}{\bibfnamefont{M.}~\bibnamefont{Schlapps}},
  \bibinfo{author}{\bibfnamefont{T.}~\bibnamefont{Lermer}},
  \bibinfo{author}{\bibfnamefont{S.}~\bibnamefont{Geissler}},
  \bibinfo{author}{\bibfnamefont{D.}~\bibnamefont{Neumaier}},
  \bibinfo{author}{\bibfnamefont{J.}~\bibnamefont{Sadowski}},
  \bibinfo{author}{\bibfnamefont{D.}~\bibnamefont{Schuh}},
  \bibinfo{author}{\bibfnamefont{W.}~\bibnamefont{Wegscheider}},
  \bibnamefont{and} \bibinfo{author}{\bibfnamefont{D.}~\bibnamefont{Weiss}},
  \bibinfo{journal}{Physical Review B} \textbf{\bibinfo{volume}{80}},
  \bibinfo{pages}{125330} (\bibinfo{year}{2009}).

\bibitem[{\citenamefont{Bernand-Mantel
  et~al.}(2009)\citenamefont{Bernand-Mantel, Seneor, Bouzehouane, Fusil,
  Deranlot, Petroff, and Fert}}]{bernand-mantel_anisotropic_2009}
\bibinfo{author}{\bibfnamefont{A.}~\bibnamefont{Bernand-Mantel}},
  \bibinfo{author}{\bibfnamefont{P.}~\bibnamefont{Seneor}},
  \bibinfo{author}{\bibfnamefont{K.}~\bibnamefont{Bouzehouane}},
  \bibinfo{author}{\bibfnamefont{S.}~\bibnamefont{Fusil}},
  \bibinfo{author}{\bibfnamefont{C.}~\bibnamefont{Deranlot}},
  \bibinfo{author}{\bibfnamefont{F.}~\bibnamefont{Petroff}}, \bibnamefont{and}
  \bibinfo{author}{\bibfnamefont{A.}~\bibnamefont{Fert}},
  \bibinfo{journal}{Nature Physics} \textbf{\bibinfo{volume}{5}},
  \bibinfo{pages}{920} (\bibinfo{year}{2009}).

\bibitem[{\citenamefont{Ciccarelli et~al.}(2012)\citenamefont{Ciccarelli,
  Z\^arbo, Irvine, Campion, Gallagher, Wunderlich, Jungwirth, and
  Ferguson}}]{ciccarelli_spin_2012}
\bibinfo{author}{\bibfnamefont{C.}~\bibnamefont{Ciccarelli}},
  \bibinfo{author}{\bibfnamefont{L.~P.} \bibnamefont{Z\^arbo}},
  \bibinfo{author}{\bibfnamefont{A.~C.} \bibnamefont{Irvine}},
  \bibinfo{author}{\bibfnamefont{R.~P.} \bibnamefont{Campion}},
  \bibinfo{author}{\bibfnamefont{B.~L.} \bibnamefont{Gallagher}},
  \bibinfo{author}{\bibfnamefont{J.}~\bibnamefont{Wunderlich}},
  \bibinfo{author}{\bibfnamefont{T.}~\bibnamefont{Jungwirth}},
  \bibnamefont{and} \bibinfo{author}{\bibfnamefont{A.~J.}
  \bibnamefont{Ferguson}}, \bibinfo{journal}{Applied Physics Letters}
  \textbf{\bibinfo{volume}{101}}, \bibinfo{pages}{122411}
  (\bibinfo{year}{2012}).

\bibitem[{\citenamefont{Ohno et~al.}(1996)\citenamefont{Ohno, Shen, Matsukura,
  Oiwa, Endo, Katsumoto, and Iye}}]{ohno_gamnas:_1996}
\bibinfo{author}{\bibfnamefont{H.}~\bibnamefont{Ohno}},
  \bibinfo{author}{\bibfnamefont{A.}~\bibnamefont{Shen}},
  \bibinfo{author}{\bibfnamefont{F.}~\bibnamefont{Matsukura}},
  \bibinfo{author}{\bibfnamefont{A.}~\bibnamefont{Oiwa}},
  \bibinfo{author}{\bibfnamefont{A.}~\bibnamefont{Endo}},
  \bibinfo{author}{\bibfnamefont{S.}~\bibnamefont{Katsumoto}},
  \bibnamefont{and} \bibinfo{author}{\bibfnamefont{Y.}~\bibnamefont{Iye}},
  \bibinfo{journal}{Applied Physics Letters} \textbf{\bibinfo{volume}{69}},
  \bibinfo{pages}{363} (\bibinfo{year}{1996}).

\bibitem[{\citenamefont{Jungwirth et~al.}(2006)\citenamefont{Jungwirth, Sinova,
  Ma\v{s}ek, Ku\v{c}era, and MacDonald}}]{jungwirth_theory_2006}
\bibinfo{author}{\bibfnamefont{T.}~\bibnamefont{Jungwirth}},
  \bibinfo{author}{\bibfnamefont{J.}~\bibnamefont{Sinova}},
  \bibinfo{author}{\bibfnamefont{J.}~\bibnamefont{Ma\v{s}ek}},
  \bibinfo{author}{\bibfnamefont{J.}~\bibnamefont{Ku\v{c}era}},
  \bibnamefont{and} \bibinfo{author}{\bibfnamefont{A.~H.}
  \bibnamefont{MacDonald}}, \bibinfo{journal}{Reviews of Modern Physics}
  \textbf{\bibinfo{volume}{78}}, \bibinfo{pages}{809} (\bibinfo{year}{2006}).

\bibitem[{\citenamefont{Gould et~al.}(2004)\citenamefont{Gould, R\"uster,
  Jungwirth, Girgis, Schott, Giraud, Brunner, Schmidt, and
  Molenkamp}}]{gould_tunneling_2004}
\bibinfo{author}{\bibfnamefont{C.}~\bibnamefont{Gould}},
  \bibinfo{author}{\bibfnamefont{C.}~\bibnamefont{R\"uster}},
  \bibinfo{author}{\bibfnamefont{T.}~\bibnamefont{Jungwirth}},
  \bibinfo{author}{\bibfnamefont{E.}~\bibnamefont{Girgis}},
  \bibinfo{author}{\bibfnamefont{G.~M.} \bibnamefont{Schott}},
  \bibinfo{author}{\bibfnamefont{R.}~\bibnamefont{Giraud}},
  \bibinfo{author}{\bibfnamefont{K.}~\bibnamefont{Brunner}},
  \bibinfo{author}{\bibfnamefont{G.}~\bibnamefont{Schmidt}}, \bibnamefont{and}
  \bibinfo{author}{\bibfnamefont{L.~W.} \bibnamefont{Molenkamp}},
  \bibinfo{journal}{Physical Review Letters} \textbf{\bibinfo{volume}{93}},
  \bibinfo{pages}{117203} (\bibinfo{year}{2004}).

\bibitem[{\citenamefont{Tongay et~al.}(2009)\citenamefont{Tongay, Hebard,
  Hikita, and Hwang}}]{tongay_magnetodielectric_2009}
\bibinfo{author}{\bibfnamefont{S.}~\bibnamefont{Tongay}},
  \bibinfo{author}{\bibfnamefont{A.~F.} \bibnamefont{Hebard}},
  \bibinfo{author}{\bibfnamefont{Y.}~\bibnamefont{Hikita}}, \bibnamefont{and}
  \bibinfo{author}{\bibfnamefont{H.~Y.} \bibnamefont{Hwang}},
  \bibinfo{journal}{Physical Review B} \textbf{\bibinfo{volume}{80}},
  \bibinfo{pages}{205324} (\bibinfo{year}{2009}).

\bibitem[{\citenamefont{Catalan}(2006)}]{catalan_magnetocapacitance_2006}
\bibinfo{author}{\bibfnamefont{G.}~\bibnamefont{Catalan}},
  \bibinfo{journal}{Applied Physics Letters} \textbf{\bibinfo{volume}{88}},
  \bibinfo{pages}{102902} (\bibinfo{year}{2006}), ISSN
  \bibinfo{issn}{0003-6951, 1077-3118}.

\bibitem[{\citenamefont{Iafrate et~al.}(1995)\citenamefont{Iafrate, Hess,
  Krieger, and Macucci}}]{Iafrate_capacitive_1995}
\bibinfo{author}{\bibfnamefont{G.~J.} \bibnamefont{Iafrate}},
  \bibinfo{author}{\bibfnamefont{K.}~\bibnamefont{Hess}},
  \bibinfo{author}{\bibfnamefont{J.~B.} \bibnamefont{Krieger}},
  \bibnamefont{and} \bibinfo{author}{\bibfnamefont{M.}~\bibnamefont{Macucci}},
  \bibinfo{journal}{Physical Review B} \textbf{\bibinfo{volume}{52}},
  \bibinfo{pages}{10737} (\bibinfo{year}{1995}).

\bibitem[{\citenamefont{Eisenstein et~al.}(1994)\citenamefont{Eisenstein,
  Pfeiffer, and West}}]{eisenstein_compressibility_1994}
\bibinfo{author}{\bibfnamefont{J.~P.} \bibnamefont{Eisenstein}},
  \bibinfo{author}{\bibfnamefont{L.~N.} \bibnamefont{Pfeiffer}},
  \bibnamefont{and} \bibinfo{author}{\bibfnamefont{K.~W.} \bibnamefont{West}},
  \bibinfo{journal}{Physical Review B} \textbf{\bibinfo{volume}{50}},
  \bibinfo{pages}{1760} (\bibinfo{year}{1994}).

\bibitem[{\citenamefont{Smith et~al.}(1986)\citenamefont{Smith, Wang, and
  Stiles}}]{smith_two-dimensional_1986}
\bibinfo{author}{\bibfnamefont{T.~P.} \bibnamefont{Smith}},
  \bibinfo{author}{\bibfnamefont{W.~I.} \bibnamefont{Wang}}, \bibnamefont{and}
  \bibinfo{author}{\bibfnamefont{P.~J.} \bibnamefont{Stiles}},
  \bibinfo{journal}{Physical Review B} \textbf{\bibinfo{volume}{34}},
  \bibinfo{pages}{2995} (\bibinfo{year}{1986}).

\bibitem[{\citenamefont{Ponomarenko et~al.}(2010)\citenamefont{Ponomarenko,
  Yang, Gorbachev, Blake, Mayorov, Novoselov, Katsnelson, and
  Geim}}]{Ponomarenko_density_2010}
\bibinfo{author}{\bibfnamefont{L.~A.} \bibnamefont{Ponomarenko}},
  \bibinfo{author}{\bibfnamefont{R.}~\bibnamefont{Yang}},
  \bibinfo{author}{\bibfnamefont{R.~V.} \bibnamefont{Gorbachev}},
  \bibinfo{author}{\bibfnamefont{P.}~\bibnamefont{Blake}},
  \bibinfo{author}{\bibfnamefont{A.~S.} \bibnamefont{Mayorov}},
  \bibinfo{author}{\bibfnamefont{K.~S.} \bibnamefont{Novoselov}},
  \bibinfo{author}{\bibfnamefont{M.~I.} \bibnamefont{Katsnelson}},
  \bibnamefont{and} \bibinfo{author}{\bibfnamefont{A.~K.} \bibnamefont{Geim}},
  \bibinfo{journal}{Physical Review Letters} \textbf{\bibinfo{volume}{105}},
  \bibinfo{pages}{136801} (\bibinfo{year}{2010}).

\bibitem[{\citenamefont{Ilani et~al.}(2006)\citenamefont{Ilani, Donev,
  Kindermann, and McEuen}}]{Ilani_measurement_2006}
\bibinfo{author}{\bibfnamefont{S.}~\bibnamefont{Ilani}},
  \bibinfo{author}{\bibfnamefont{L.~a.~K.} \bibnamefont{Donev}},
  \bibinfo{author}{\bibfnamefont{M.}~\bibnamefont{Kindermann}},
  \bibnamefont{and} \bibinfo{author}{\bibfnamefont{P.~L.}
  \bibnamefont{McEuen}}, \bibinfo{journal}{Nature Physics}
  \textbf{\bibinfo{volume}{2}}, \bibinfo{pages}{687} (\bibinfo{year}{2006}).

\bibitem[{\citenamefont{Hansen et~al.}(1989)\citenamefont{Hansen, Smith, Lee,
  Brum, Knoedler, Hong, and Kern}}]{hansen_zeeman_1989}
\bibinfo{author}{\bibfnamefont{W.}~\bibnamefont{Hansen}},
  \bibinfo{author}{\bibfnamefont{T.~P.} \bibnamefont{Smith}},
  \bibinfo{author}{\bibfnamefont{K.~Y.} \bibnamefont{Lee}},
  \bibinfo{author}{\bibfnamefont{J.~A.} \bibnamefont{Brum}},
  \bibinfo{author}{\bibfnamefont{C.~M.} \bibnamefont{Knoedler}},
  \bibinfo{author}{\bibfnamefont{J.~M.} \bibnamefont{Hong}}, \bibnamefont{and}
  \bibinfo{author}{\bibfnamefont{D.~P.} \bibnamefont{Kern}},
  \bibinfo{journal}{Physical Review Letters} \textbf{\bibinfo{volume}{62}},
  \bibinfo{pages}{2168} (\bibinfo{year}{1989}).

\bibitem[{\citenamefont{Ashoori et~al.}(1992)\citenamefont{Ashoori, Stormer,
  Weiner, Pfeiffer, Pearton, Baldwin, and West}}]{ashoori_single-electron_1992}
\bibinfo{author}{\bibfnamefont{R.~C.} \bibnamefont{Ashoori}},
  \bibinfo{author}{\bibfnamefont{H.~L.} \bibnamefont{Stormer}},
  \bibinfo{author}{\bibfnamefont{J.~S.} \bibnamefont{Weiner}},
  \bibinfo{author}{\bibfnamefont{L.~N.} \bibnamefont{Pfeiffer}},
  \bibinfo{author}{\bibfnamefont{S.~J.} \bibnamefont{Pearton}},
  \bibinfo{author}{\bibfnamefont{K.~W.} \bibnamefont{Baldwin}},
  \bibnamefont{and} \bibinfo{author}{\bibfnamefont{K.~W.} \bibnamefont{West}},
  \bibinfo{journal}{Physical Review Letters} \textbf{\bibinfo{volume}{68}},
  \bibinfo{pages}{3088} (\bibinfo{year}{1992}).

\bibitem[{\citenamefont{B\"uttiker}(1993)}]{buttiker_capacitance_1993}
\bibinfo{author}{\bibfnamefont{M.}~\bibnamefont{B\"uttiker}},
  \bibinfo{journal}{Journal of Physics: Condensed Matter}
  \textbf{\bibinfo{volume}{5}}, \bibinfo{pages}{9361} (\bibinfo{year}{1993}).

\bibitem[{\citenamefont{Kopp and Mannhart}(2009)}]{kopp_calculation_2009}
\bibinfo{author}{\bibfnamefont{T.}~\bibnamefont{Kopp}} \bibnamefont{and}
  \bibinfo{author}{\bibfnamefont{J.}~\bibnamefont{Mannhart}},
  \bibinfo{journal}{Journal of Applied Physics} \textbf{\bibinfo{volume}{106}},
  \bibinfo{pages}{064504} (\bibinfo{year}{2009}).

\bibitem[{\citenamefont{Neumaier et~al.}(2009)\citenamefont{Neumaier, Turek,
  Wurstbauer, Vogl, Utz, Wegscheider, and
  Weiss}}]{neumaier_all-electrical_2009}
\bibinfo{author}{\bibfnamefont{D.}~\bibnamefont{Neumaier}},
  \bibinfo{author}{\bibfnamefont{M.}~\bibnamefont{Turek}},
  \bibinfo{author}{\bibfnamefont{U.}~\bibnamefont{Wurstbauer}},
  \bibinfo{author}{\bibfnamefont{A.}~\bibnamefont{Vogl}},
  \bibinfo{author}{\bibfnamefont{M.}~\bibnamefont{Utz}},
  \bibinfo{author}{\bibfnamefont{W.}~\bibnamefont{Wegscheider}},
  \bibnamefont{and} \bibinfo{author}{\bibfnamefont{D.}~\bibnamefont{Weiss}},
  \bibinfo{journal}{Physical Review Letters} \textbf{\bibinfo{volume}{103}},
  \bibinfo{pages}{087203} (\bibinfo{year}{2009}).

\bibitem[{\citenamefont{Shick et~al.}(2006)\citenamefont{Shick, M\'{a}ca,
  Ma\v{s}ek, and Jungwirth}}]{shick_prospect_2006}
\bibinfo{author}{\bibfnamefont{A.~B.} \bibnamefont{Shick}},
  \bibinfo{author}{\bibfnamefont{F.}~\bibnamefont{M\'{a}ca}},
  \bibinfo{author}{\bibfnamefont{J.}~\bibnamefont{Ma\v{s}ek}},
  \bibnamefont{and}
  \bibinfo{author}{\bibfnamefont{T.}~\bibnamefont{Jungwirth}},
  \bibinfo{journal}{Physical Review B} \textbf{\bibinfo{volume}{73}},
  \bibinfo{pages}{024418} (\bibinfo{year}{2006}).

\bibitem[{\citenamefont{Park et~al.}(2008)\citenamefont{Park, Wunderlich,
  Williams, Joo, Jung, Shin, Olejn\'ik, Shick, and
  Jungwirth}}]{park_tunneling_2008}
\bibinfo{author}{\bibfnamefont{B.~G.} \bibnamefont{Park}},
  \bibinfo{author}{\bibfnamefont{J.}~\bibnamefont{Wunderlich}},
  \bibinfo{author}{\bibfnamefont{D.~A.} \bibnamefont{Williams}},
  \bibinfo{author}{\bibfnamefont{S.~J.} \bibnamefont{Joo}},
  \bibinfo{author}{\bibfnamefont{K.~Y.} \bibnamefont{Jung}},
  \bibinfo{author}{\bibfnamefont{K.~H.} \bibnamefont{Shin}},
  \bibinfo{author}{\bibfnamefont{K.}~\bibnamefont{Olejn\'ik}},
  \bibinfo{author}{\bibfnamefont{A.~B.} \bibnamefont{Shick}}, \bibnamefont{and}
  \bibinfo{author}{\bibfnamefont{T.}~\bibnamefont{Jungwirth}},
  \bibinfo{journal}{Physical Review Letters} \textbf{\bibinfo{volume}{100}},
  \bibinfo{pages}{087204} (\bibinfo{year}{2008}).

\end{thebibliography}
\bibliographystyle{apsrev}

\end{document}